# Decentralized Inter-User Interference Suppression in Body Sensor Networks with Non-cooperative Game

Guowei Wu, Jiankang Ren, Feng Xia, Lin Yao, and Zichuan Xu
School of Software, Dalian University of Technology, Dalian 116620, China
E-mail: wgwdut@dlut.edu.cn; rjk.dlut@gmail.com; f.xia@ieee.org; yaolin@dlut.edu.cn; eggerxug@gmail.com

*Abstract*—Body Sensor Networks (BSNs) provide continuous health monitoring and analysis of physiological parameters. A high degree of Quality-of-Service (QoS) for BSN is extremely required. Inter-user interference is introduced by the simultaneous communication of BSNs congregating in the same area. In this paper, a decentralized inter-user interference suppression algorithm for BSN, namely DISG, is proposed. Each BSN measures the SINR from other BSNs and then adaptively selects the suitable channel and transmission power. By utilizing non-cooperative game theory and no regret learning algorithm, DISG provides an adaptive inter-user interference suppression strategy. The correctness and effectiveness of DISG is theoretically proved, and the experimental results show that DISG can reduce the effect of inter-user interference effectively.

*Keywords-Body sensor networks; QoS; inter-user interference; game theory; no regret learning algorithm*

## I. INTRODUCTION

With recent advances in intelligent (bio-) medical sensors, low-power integrated circuits and the rapid development of wireless communication, Body Sensor Networks (BSNs) have been applied in remote health monitoring and patient care [1] [2] [3]. By outfitting patients with wireless wearable or implanted vital sign sensors, detailed real-time data on physiological status can be continuously sampled, such as ECG, SpO2, blood pressure, etc. Although BSNs share many of the same challenges with general wireless sensor networks (WSNs), many BSN-specific challenges have emerged.

BSN needs high degrees of reliability for personal real-time health monitoring [4], however, BSN has fewer and smaller nodes relative to a conventional WSN. Smaller nodes imply smaller batteries, creating strict tradeoffs between the energy consumption, storage, and communication resources. Usually, as wireless sensor network, when users gather in the same place and their sensors try to transmit data simultaneously, the transmissions of sensors in different users will interfere with each other. Inter-user interference causes unreliable critical data transmission and high bite error rate, hence, the critical data can't be sent to control nodes in time. Consequently, the doctor may give wrong diagnosis and the patient may be delayed to cure and even die. Moreover, the packet loss results in data retransmission. Due to the characteristics of energy-sensitiveness and resource-limitedness, numerous data retransmissions consume much energy and thus may cause the failure of biosensors. Additionally, there is almost no underlying inter-user interference suppression infrastructure in BSN. Thus, providing a decentralized inter-user interference suppression algorithm for BSN is becoming urgent.

In this paper, a decentralized inter-user interference suppression algorithm for BSN, namely DISG, is proposed. In DISG, each BSN measures the SINR from other BSNs and then adaptively selects the suitable channel and transmission power to reduce the effect of inter-user interference. Through non-cooperative game theory, DISG provides an adaptive inter-user interference suppression strategy by utilizing the no regret learning algorithm.

The rest of the paper is organized as follows. Section II introduces related work. In Section III the application scenario of DISG algorithm is given. Section IV gives a detailed description of the game theoretic framework in DISG. The performance of DISG is evaluated in Section V. Finally, we conclude the paper in Section VI.

## II. RELATED WORK

The growing interest in BSN and the continual emergence of new techniques have inspired some efforts to research QoS of BSN. Extensive research has been focusing on the interference suppression in WSN. Fasano [5] proposes a novel consensus algorithm based on [6] to suppress both noise and interference. In [7], the authors propose a cluster-based WSN MAC protocol CBPMAC/TFC. Wu and Biswas [8] present a self-reorganizing slot allocation mechanism for TDMA-based MAC in multi-cluster sensor networks. The issue of interference in WSN is similar to the problem of inter-user interference in BSN. However, there is a significant gap between the conventional WSN systems due to the characteristics of BSN, especially the mobility of BSN, and thus none of these interference suppression approaches in WSN can be directly applied in BSN.

Currently, the reliability research of BSN mainly focuses on the reliability of communications in intra-BSN. In [9] [10], a novel QoS cross-layer scheduling mechanism based on fuzzy-logic rules is proposed. Braem *et al.* [11] [12] propose an active replication algorithm for the reliability of communication in multi-hop BSN. In [13], a statistical bandwidth strategy, namely BodyQoS, is proposed to guarantee reliable data communication. Nonetheless, all of them neglect the inter-user interference. Natarajan *et al.* [14] highlights the existence of the inter-user interference effect in BSN from the perspective of network architectures, but the issue of inter-user interference is not investigated. In [15], the authors conduct a preliminary investigation of the impact of inter-user interference, and implement an instance with a fixed WSN infrastructure to reduce interference between users. However, that paper does not conduct a comprehensive study of inter-user interference. In addition, the fixed WSN infrastructure is only suitable for indoor environments.

Although the existing researches play important roles to improve reliability of BSN, the inter-user interference is still a challenging issue in BSN. In this paper, a decentralized inter-user interference suppression algorithm is demonstrated. The major differences between this work and the aforementioned researches include the following aspects:

- Non-cooperative game theory is utilized, thus DISG is distributed, non-cooperative and self-organizing. In this architecture, each BSN measures the SINR from other BSNs and then adaptively selects the suitable channel and transmission power without the need of the communication with each other.
- Compared with [15], in DISG, the deployment of the fixed WSN infrastructure is not required. Therefore the system with DISG is more flexible. Moreover, DISG suppresses the inter-user interference not only through the channel selection, but also through the transmission power adjustment.

## III. SYSTEM MODEL

In this section, we describe the application scenario of the DISG algorithm. The scenario where the BSN is formed by a collection of biosensors, control nodes and a base station is considered in this paper. Fig.1 shows the system scenario. In general, the biosensors are wireless wearable or implanted vital sign sensors. It consists of a processor, memory, transceiver, sensor and a power unit. Each node is typically capable of sensing, processing, storing and transmitting the data. Through the base station, sampled data are periodically sent to the medical server in hospital by the control node, where they are stored for further processing [16].

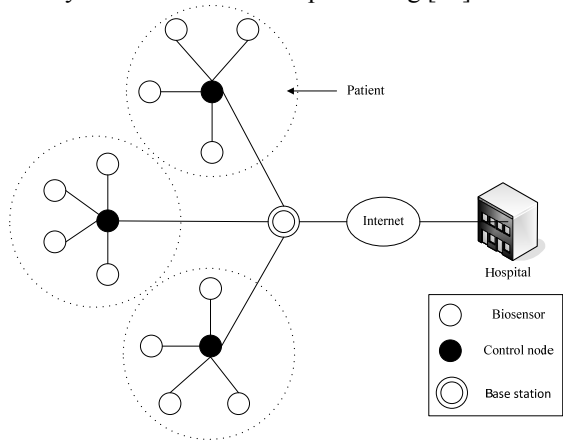

Figure 1. BSN application scenario.

In the scenario illustrated in Fig.1, the transmissions of sensors in different users will interfere with each other, when they operate in the same vicinity and communicate concurrently. Such inter-user interference will be increased in some architecture predisposing some of sensors to transmit with high power. Additionally, some extraneous interference in the same band can also increase the inter-user interference [14]. DISG algorithm reduces the effect of inter-user interference in the network to guarantee reliability performance by modeling the channel selection as a non-cooperative game to decentralize the inter-user interference suppression. In the game, each user measures the SINR from other users and then adaptively selects the suitable channel and transmission power to suppress the inter-user interference by utilizing no regret learning algorithm.

## IV. GAME THEORETIC FRAMEWORK

### A. Inter-user Interference Suppression Game Mode

In order to reduce the effect of inter-user interference in the network, we use the technique of channel selection [15] and model the channel selection of users as a non-cooperative game. In this game, each user is a player and denoted by a set $B = \{b_i, i = 1,...,M\}$, and $M$ is the number of users. Let $C = \{c_j, j = 1,...,N\}$ represents the set of all non-overlapping frequency channels that can be used by the users, where $N$ is the number of the available non-overlapping channels. The transmission power of $b_i$ is $p_i \in [p_{\min}, p_{\max}]$ due to the characteristic of the specific physiological sensors. The received power of $b_i$ is expressed as $G_i p_i$, where $G_i > 0$ is the link gain of $b_i$ and can be calculated as follows:

$$G_i = \frac{S_i}{d_i^{\delta}}. \qquad (1)$$

where $S_i$ is the attenuation coefficient, $d_i$ is the maximum distance between the biosensor node and the control node in $b_i$. $\delta$ is a constant parameter modeling the shadowing effect. Based on [17], the signal to interference plus noise ratio (SINR) of $b_i$ on the channel $c_j$ is calculated as follows:

$$\gamma_i^j = \frac{G_i p_i^j}{\sum_{k=1, k \neq i}^{M} \frac{S_k p_k^j}{(v_k^i)^{\delta}} + \eta_i}. \qquad (2)$$

where $p_i^j$ is the transmission power of $b_i$ on channel $c_j$, $v_k^i$ is the distance between user $b_k$ and user $b_i$, and $\eta_i$ is the background noise power. In the game, each user $b_i$ is selfish and only cares about itself to close to the ideal SINR threshold $\gamma_i^0$ to ensure specific QoS requirement at an optimal power level on an optimal channel, but whether other users meet their QoS requirements is irrelevant. Obviously, this framework of the game is suitable for solving the problem of the inter-user interference.

**Definition 1** Non-cooperative inter-user interference suppression game. The non-cooperative game is represented by the following tuple $\langle B, \Lambda, U \rangle$.

$\Lambda$ is the action space of the system, where $\Lambda = \Lambda_1 \times ... \times \Lambda_M$. The strategy space of user $b_i$ is defined as $\Lambda_i = \{\sigma_i := \{\rho_i^j p_i^j\} \mid j \in \{1,...,N\}\}$, where we define the mixed strategy as $\rho_i^j \in \{0,1\}$, $\sum_{j=1}^{N} \rho_i^j = 1$, and $\rho_i^j = 1$ indicates that $b_i$ selects channel $c_j$, which means in this distribution, only the pure strategy $\rho_i^j$ is assumed to be selected with a probability of 1. $p_i^j$ represents the transmission power of $b_i$ on channel $c_j$ and $p_i^j \in [p_{\min}, p_{\max}]$.

U is the utility set of the system, where $U = \{u_1,...,u_M\}$. The utility of user $b_i$ is denoted as $u_i(\sigma_i, \sigma_{-i})$, which is a function of all users' actions. Each user selects its own action $\sigma_i$ to minimize the utility function. $\sigma_{-i}$ is the action taken by users except $b_i$.

There is a trade-off between SINR and power utilization due to the energy-limitedness of biosensor, thus, similar to [18] [19], the utility function that considers SINR and power trade-offs is expressed as:

$$u_i(\sigma_i, \sigma_{-i}) = \tau_i(\gamma_i^0 - \gamma_i^j)^2 + \xi_i p_i^j, \rho_i^j = 1. \quad (3)$$

where $\tau_i$ and $\xi_i$ are non-negative weighting factors. Different levels of emphasis on SINR and power are obtained by adjusting the weighting factors. Based on the equation (2), we can get:

$$u_i(\sigma_i, \sigma_{-i}) = \tau_i \left( \gamma_i^0 - \frac{G_i p_i^j}{\sum_{k=1, k \neq i, \rho_k^j = 1}^{M} \frac{S_k p_k^j}{(\upsilon_k^i)^\delta} + \eta_i} \right)^2 + \xi_i p_i^{j_i}, \rho_i^j = 1. \quad (4)$$

In the game, the goal of user $b_i$ is to attain high SINR and minimize power utilization over the channel that it chooses. It is worthwhile to note that there is no message exchange in the game. The utility distribution of other users $\sigma_{-i}$ is not available for user $b_i$. However, $b_i$ can get the aggregate response $I_i^j(\sigma_{-i})$. To perform equation (5), user $b_i$ needs to observe the local information $I_i = \{I_i^j(\sigma_{-i}), \forall c_j\}$. Based on the information, the following best response is adopted by every user in the system:

$$\pi_i(I_i) = \arg\min_{\sigma_i \in \Lambda_i} u_i(\sigma_i, I_i) \quad \forall b_i \in B. \quad (5)$$

where $\pi_i$ represents the strategy for the selection of the channel and the transmission power. The solution to the problem in equation (5) will lead to a Nash Equilibrium (NE).

*B. Nash Equilibrium in Game*

In order to analyze the outcome of the game, we will give how the game is played by rational players using the concept of NE, which states that in the equilibrium every player will select a utility-minimizing strategy given the strategies of other players.

**Definition 2** Nash Equilibrium:

$$\rho^* p^* = \left[ \rho_1^{j_1^*} \left( p_1^{j_1^*} \right)^*, \rho_2^{j_2^*} \left( p_2^{j_2^*} \right)^*, ..., \rho_N^{j_N^*} \left( p_N^{j_N^*} \right)^* \right]. \quad (6)$$

is defined as:

$$u_i \left( \rho_i^{j_i^*} \left( p_i^{j_i^*} \right)^*, \rho_{-i}^j p_{-i}^j \right) \leq u_i \left( \rho_i' p_i', \rho_{-i}^j p_{-i}^j \right), \forall b_i \in B, \forall \rho_i' p_i' \in \Lambda_i. \quad (7)$$

i.e., given other players' actions, no user can reduce its utility alone by changing its own action.

**Theorem 1** For a specific channel $c_j$, with all other players' transmission powers fixed on the channel, user $b_i$ can get the best transmission power on the channel $c_j$ $(p_i^{j_i})^* = \arg\min_{p_i^{j_i} \in p_i} u_i(p_i^{j_i}, p_{-i}^{j_i})$ in the condition of $\tau_i / \xi_i \geq 2 p_{\max} / (\gamma_i^0)^2$ and $I_i^{j_i}(p_{-i}^{j_i}) < \tau_i \gamma_i^0 G_i / \xi_i$.

**Proof:** Let $\sum_{k=1, k \neq i, j_k \neq j_i}^{M} S_k p_k^j / (\upsilon_k^i)^\delta + \eta_i = I_i^{j_i}(p_{-i}^{j_i})$, so equation (4) can be written as:

$$u_i(p_i^{j_i}, p_{-i}^{j_i}) = \tau_i(\gamma_i^0 - \frac{G_i p_i^{j_i}}{I_i^{j_i}(p_{-i}^{j_i})})^2 + \xi_i p_i^{j_i}, \rho_i^{j_i} = 1. \quad (8)$$

Take the derivative with respect to $p_i^{j_i}$ and equate it to zero for $p_i^{j_i}$ to be a NE, then we can get:

$$p_i^{j_i} = \frac{\gamma_i^0 I_i^{j_i}(p_{-i}^{j_i})}{G_i} - \frac{\xi_i \left( I_i^{j_i}(p_{-i}^{j_i}) \right)^2}{2\tau_i G_i^2}. \quad (9)$$

Due to $p_i^{j_i} \geq 0$, we can get:

$$I_i^{j_i}(p_{-i}^{j_i}) \leq \frac{2\tau_i \gamma_i^0}{\xi_i}. \quad (10)$$

For given $\tau_i$, $\xi_i$, $\gamma_i^0$, $G_i$ and $p_i^{j_i}$, let the quadratic equation (9) in $I_i^{j_i}(p_{-i}^{j_i})$ have a real solution. Because the transmission powers is $p_i^j \leq p_{\max}$, $\tau_i / \xi_i$ must satisfy:

$$\frac{\tau_i}{\xi_i} \geq \frac{2p_{\max}}{\left(\gamma_i^0\right)^2}. \qquad (11)$$

According to [17], in order to make the game converge to a unique fixed point, a fixed point from the equation (9) should satisfy the properties: 1) positivity; 2) monotonicity; 3) scalability. Based on [18], we can get

$$I_i^{j_i}(p_{-i}^{j_i}) < \frac{\tau_i \gamma_i^0 G_i}{\xi_i}. \qquad (12)$$

Based on equations (10), (11) and (12), the non-cooperative game has a unique NE for the transmission power $(p_i^{j_i})^*$ on a certain channel $c_j$ in the condition of $\tau_i/\xi_i \geq 2p_{\max}/(\gamma_i^0)^2$ and $I_i^{j_i}(p_{-i}^{j_i}) < \tau_i \gamma_i^0 G_i/\xi_i$. ∎

**Definition 3** A probability distribution $\omega$ is a correlated equilibrium (CE) over the set of user strategies, if and only if, the following inequality is satisfied:

$$\sum_{\rho_{-i} \in \Lambda_i} \omega(\rho_i, \rho_{-i}) \left[ u_i(\rho_i', \rho_{-i}) - u_i(\rho_i, \rho_{-i}) \right] \geq 0. \qquad (13)$$

for all $b_i \in B$, and $\rho_i \in \Lambda_i$ and $\rho_{-i} \in \Lambda_{-i}$, and a CE is an NE if and only if it is a product measure.

**Theorem 2** Each user $b_i$ converges to a NE to attain the best channel $c_{j_i^*}$ (i.e., $\rho_i^{j_i} = 1$) with the no regret learning approach.

**Proof**: Based on the definition 3, we can find the set of correlated equilibrium is nonempty, convex and closed in the finite game. Thus, according to [20], when players select their strategy randomly, there exists a no-regret strategy such that if every player follows such a strategy, then the empirical frequencies of play converge to the set of correlated equilibrium. In addition, this correlated equilibrium is NE, because the correlated equilibrium corresponds to the special case in which $\omega(\rho_i, \rho_{-i})$ is a product of each player's probability for different actions, i.e., the play of the different players is independent. Thus, based on the NE converged, each user can attain the best channel by no regret learning approach. ∎

**Theorem 3** Nash Equilibrium exists in the non-cooperative inter-user interference suppression game.

**Proof:** For each player $b_i \in B$, it can attain the best channel $c_{j_i^*}$ (i.e., $\rho_i^{j^*} = 1$) with the no regret learning approach based on the Theorem 2, and select the best transmission power $p_i^{j_i^*}$ on the channel $c_{j_i^*}$ according to Theorem 1 to minimize utility. Therefore, there exists a NE $\rho^* p^* = \left[ \rho_1^{j_1^*}\left(p_1^{j_1^*}\right)^*, \rho_2^{j_2^*}\left(p_2^{j_2^*}\right)^*, ..., \rho_N^{j_N^*}\left(p_N^{j_N^*}\right)^* \right]$ in the game. ∎

### C. No Regret Learning Strategy

The no-regret learning algorithm [21] is utilized to get the NE in the game. In the game, each user selects its action based on the history of actions of every user. User $b_i$ determines the probability distribution of strategy $\sigma_i$ based on the average regret at iteration period $\varphi \in \{1...t\}$. For two distinct strategies $\left({}^\varphi\sigma_i, {}^\varphi\sigma_{-i}\right)$ and $\left(\sigma_i\left(\left(p_i^j\right)^*\right), {}^\varphi\sigma_{-i}\right)$, the historical cumulative regret of user $b_i$ from $\left({}^\varphi\sigma_i, {}^\varphi\sigma_{-i}\right)$ to $\left(\sigma_i\left(\left(p_i^j\right)^*\right), {}^\varphi\sigma_{-i}\right)$ up to iteration $t$ is:

$${}^t D_i\left(\sigma_i\left(p_i^j\right)^*\right) = \max\left\{\frac{1}{t}\sum_{\varphi=1}^{t}\left[u_i\left({}^\varphi\sigma\right) - u_i\left(\sigma_i\left(\left(p_i^j\right)^*\right), {}^\varphi\sigma_{-i}\right)\right], 0\right\}. \qquad (14)$$

Based on (14), the probability $\omega_i$ of the selection of action $\sigma_i\left(p_i^j\right)^*$ of user $b_i$ at iteration $t+1$ is calculated as follows:

$${}^{t+1}\omega_i\left(\sigma_i\left(p_i^j\right)^*\right) = \frac{{}^t D_i\left(\sigma_i\left(\left(p_i^j\right)^*\right)\right)}{\sum_{\sigma_i(p_i^{j'}) \in \Lambda_i} {}^t D_i\left(\sigma_i\left(\left(p_i^{j'}\right)^*\right)\right)}. \qquad (15)$$

Each user $b_i$ repeats calculations based on equations (14), (15), and then it can determine its strategy at time step $t+1$ as:

$${}^{t+1}\rho_i\left(\sigma_i\left(\left(p_i^j\right)^*\right)\right) = \begin{cases} 1 & \text{if } {}^{t+1}\omega_i\left(\sigma_i\left(\left(p_i^j\right)^*\right)\right) = \max\left\{{}^{t+1}\omega_i\right\} \\ 0 & \text{otherwise} \end{cases}. \qquad (16)$$

Fig.2 illustrates the learning strategy with no-regret learning algorithm. At the initial step, each user $b_i$ randomly selects its channel $c_{j_i}$ with equal probability. The optimal transmission power $p_i^{j_i}$ and utility $u_i(\sigma_i, \sigma_{-i})$ on channel $c_{j_i}$ will be calculated. Based on $p_i^{j_i}$ and $u_i(\sigma_i, \sigma_{-i})$, the probability of $b_i$ for the current strategy is got. If there exists a channel that has not been selected, then user $b_i$ repeats the step of random channel selection with equal probability. When the maximum $\omega_i\left(\sigma_i\left(p_i^j\right)^*\right)$ in the probability distribution $\omega_i$ has exceeded the threshold $\phi$, user $b_i$ will get the solution $\rho_i^{j_i^*}\left(p_i^{j_i^*}\right)^*$, otherwise it will select a channel based on the probability distribution $\omega_i$ and repeats its learning process.

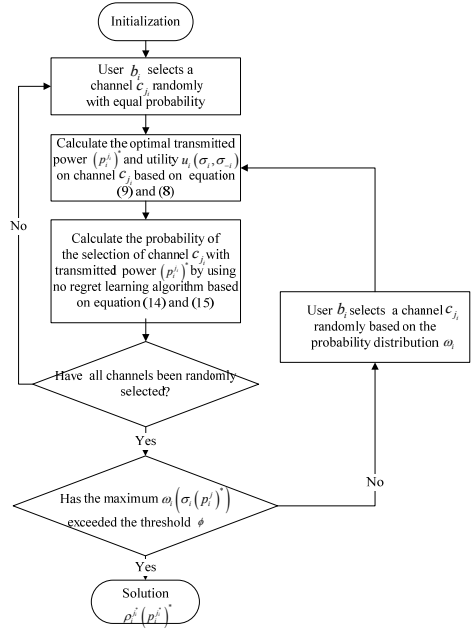

Figure 2. Learning strategy in the game.

## V. PERFORMANCE ANALYSIS

In order to evaluate DISG, we extend the Castalia simulator [22] to support the simulation of inter-user interference. Castalia is an open source, discrete event-driven simulator designed specifically for WSN and BAN based on OMNeT++ [23].

In the experiment, there exist five BSNs deployed randomly in a 10m×10m field, and each BSN simulates a typical on-body sensor network, in which sensors measure a person's physiological parameters. Each BSN has a control node and three biosensors. All sensor nodes adopt Castalia standard CC2420 IEEE802.15.4 radios. Five non-overlapping frequency channels can be used by each BSN in the experiment. Table I describes the detailed simulation parameters.

TABLE I. SIMULATION PARAMETERS

| Parameters | Value |
|---|---|
| Number of BSNs | 5 |
| Number of sensor types | 3 { ECG, SpO2, Temperature } |
| Number of non-overlapping frequency channels | 5 {11, 12, 13, 14, 15} |
| Transmission power set (mW) | {29.04, 32.67, 36.3, 42.24, 46.2, 50.69, 55.18, 57.42} |
| Wireless channel model | Log shadowing wireless model |
| Path loss exponent | 2.4 |
| Collision model | Additive interference model |
| Physical and MAC layer | IEEE 802.15.4 standard |

In the whole experiment, three metrics are used for performance evaluation: the transmission power of biosensor of each BSN, the channel used by each BSN and the average SINR of the system.

Fig.3 presents the channel selection of each BSN with DISG. At the beginning of the experiment, all BSNs select their channels randomly with equal probability. Then they select channels based on the probability distribution got by the no regret learning. At last, the experiment converges to the point where all BSNs operate on different frequency channels to reduce the inter-user interference effect.

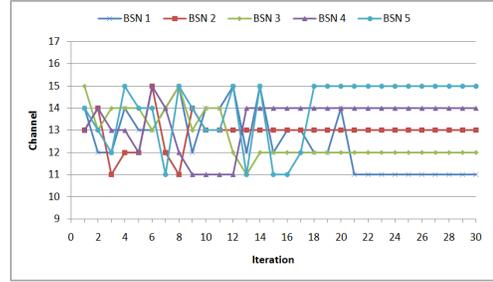

Figure 3. Channel selection of each BSN with DISG.

The transmission power selection of each BSN with DISG is presented in Fig.4. We can find that each BSN selects the suitable transmission power at each iteration to get the lowest utility, and then calculates the regret to get the probability distribution of the action. When the channel selections of all BSNs converge, each BSN will select the suitable transmission power over the optimal channel.

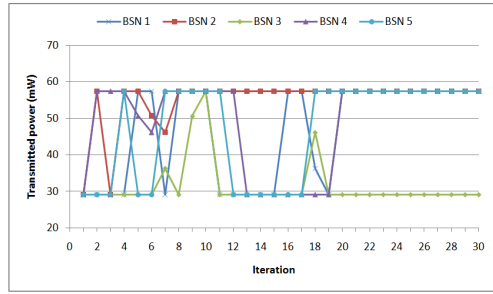

Figure 4. Transmission power selection of each BSN with DISG.

As can be seen from Fig.5, the average SINR of all BSNs in the system with DISG is obviously higher than the system without DISG in the entire experimental process. In the phase of channel random selection with equal probability, DISG considers the trade-off between SINR and power utilization, thus the average SINR is higher owing to the adaptive selection of transmission power. In addition, the no regret learning can further improve the average SINR by the channel selection based on the probability distribution. At the end of the experiment, the average SINR of the system with DISG is almost 90dB and relatively stable compared with the system without operating DISG. Therefore, DISG can reduce the inter-user interference effectively.

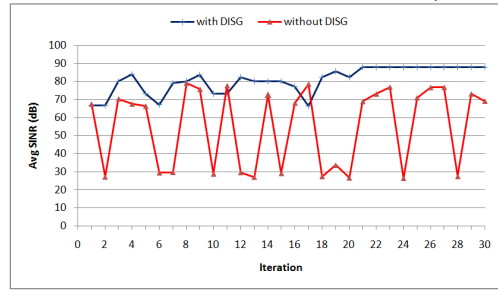

Figure 5. Average SINR of the system.

## VI. Conclusion

In this paper, a decentralized inter-user interference suppression algorithm with non-cooperative game for BSN (DISG) has been presented. DISG can effectively reduce the inter-user interference effect by utilizing no regret learning algorithm. We theoretically prove the correctness and effectiveness of DISG. Simulation results show that DISG achieves high performance. The primary contributions of this paper are summarized as follows:

(1) A non-cooperative game theoretic framework is adopted, in which the control node of each user measures the SINR from other users and then adaptively selects the best channel and transmission power to suppress the inter-user interference.

(2) The trade-off between SINR and power utilization is considered in the inter-user interference suppression game.

(3) Compared with the existing infrastructure-based inter-user interference suppression solution, the deployment of the fixed WSN infrastructure is not required in DISG.

## Acknowledgment

This work was partially supported by the National Natural Science Foundation of China under Grant No. 60703101, No. 60903153 and the Fundamental Research Funds for the Central Universities.